\documentclass{PoS}

\usepackage{amsmath,amssymb}
\usepackage{bm}
\usepackage[utf8]{inputenc}

\title{Radiative decays of light-quark mesons to a pion revisited in the covariant oscillator
 quark model}

\ShortTitle{Radiative decays of light-quark mesons to a pion revisited}

\author{\speaker{Kenji Yamada}\\
        Department of Science and Manufacturing Technology, Junior College Funabashi Campus,
         Nihon University\\
        E-mail: \email{yamada.kenji@nihon-u.ac.jp}}

\author{Tomohito Maeda\\
        Department of Science and Manufacturing Technology, Junior College Funabashi Campus,
         Nihon University\\
        E-mail: \email{maeda.tomohito@nihon-u.ac.jp}}

\abstract{
The present talk consists of two parts. The first one is allocated for consideration of
 the dimension of bilocal meson fields for quark-antiquark meson systems, which have
 been so far treated as the bosonic fields independent of constituent quark flavors,
 in the covariant oscillator quark model. Revisiting the electromagnetic currents of
 quark-antiquark meson systems, we show that the bilocal meson fields should be bosonic
 for light-quark meson systems, while fermionic for heavy-light and heavy-heavy meson
 systems.
 In the second part we apply the effective electromagnetic interactions of
 meson systems derived in the first part to radiative decay processes for
 the excited states of light-quark mesons. The calculated results for the
 $\pi \gamma$ decay widths of the $a_1(1260)$, $a_2(1320)$, $b_1(1235)$
 and $\pi_2(1670)$ mesons are in fair agreement with experiment,
 except for the $b_1(1235)$ meson.
 As for the $\rho(770)^0 \gamma$ decay width of the $f_1(1285)$
 the present model strongly supports the experimental results in 1995 and 2016,
 respectively, of VES and CLAS Collaborations, not the average value
 by the Particle Data Group.
}

\FullConference{XVII International Conference on Hadron Spectroscopy and Structure - Hadron2017\\
		25-29 September, 2017\\
		University of Salamanca, Salamanca, Spain}

\begin{document}

\section{Introduction}

In dealing with radiative decays some typical approximations, such as long wavelength
 and nonrelativistic approximations, which are not always justified, are usually used.
 For instance, the recoil effect of final-state mesons is
 neglected, though the momenta of those mesons are often comparable to their masses,
 especially in light-quark meson sectors.
 There is also ambiguity associated with a choice of the relativistic or nonrelativistic
 phase space in the nonrelativistic quark model. When the decays become relativistic,
 no one knows the rigorous way of deriving the relationship between nonrelativistic
 decay amplitudes and relativistic decay widths.

In the covariant oscillator quark model (COQM), on the other hand, hadrons themselves
 and their interactions are formulated in a manifestly covariant way. Since both the
 interaction operator and wave functions are relativistic, the gauge invariance is
 preserved and the conserved electromagnetic currents of hadrons are given explicitly
 in terms of the constituent quark variables \cite{IYO}.

\section{The covariant oscillator quark model}

\textit{Basic framework of the COQM.}
 In the COQM quark-antiquark meson systems are described by the bilocal field
\begin{equation}
	\Psi(x_{1},x_{2})_\alpha{}^\beta = \Psi(X,x)_\alpha{}^\beta,
\end{equation}
where $x_{1}^{\mu}~(x_{2}^{\mu})$ is the space-time coordinate,
 $\alpha~(\beta)$ the Dirac spinor index of the constituent quark (antiquark),
 and the center-of-mass and relative coordinates are defined, respectively, by
\begin{equation}
	X^{\mu}=\frac{m_{1}x_{1}^{\mu}+m_{2}x_{2}^{\mu}}{m_{1}+m_{2}},~~
	x^{\mu}=x_{1}^{\mu}-x_{2}^{\mu}
\end{equation}
with the constituent quark (antiquark) mass $m_{1}~(m_{2})$.
The bilocal meson field is required to satisfy the Klein-Gordon-type equation
\begin{equation}
	\left(-\frac{\partial^2}{\partial X_{\mu}\partial X^{\mu}}-\mathcal{M}^2(x)\right) 
	\Psi(X, x)_{\alpha}{}^{\beta}=0
\end{equation}
with the squared-mass operator, in the pure confining force limit,
\begin{equation}
	{\cal M}^2(x) = 2(m_1+m_2) \left(\frac{1}{2\mu}\frac{\partial^2}
	{\partial x_{\mu}\partial x^{\mu}}+U(x) \right),~~
	U(x)=-\frac{1}{2}Kx_\mu x^\mu+\rm const.
\end{equation}
where $\mu = {m_{1}m_{2}}/({m_{1}+m_{2}})$ is the reduced mass and $K$ is
 the spring constant. A solution of this equation can be written as
\begin{equation}
	\Psi (X, x)_{\alpha}{}^{\beta} = N e^{\mp i P_{\mu}X^{\mu}}
	\Phi (v, x)_{\alpha}{}^{\beta},~~v^{\mu} = P^{\mu}/M,
\end{equation}
where $N$ is the normalization constant for the plane wave, $P^{\mu}$ and $M$
 are the center-of-mass momentum and mass, respectively, of the whole meson system,
 and $\Phi (v, x)_{\alpha}{}^{\beta}$ is the internal wave function which is given
 by the direct product of eigenfunctions of the squared-mass operator and
 the Bargmann-Wigner spinor functions, defined by the direct tensor product of
 respective Dirac spinors, with the meson four velocity $v^{\mu}$, for the constituent
 quark and antiquark.

\textit{Key features of the COQM.}
 In order to freeze the redundant freedom of relative time for the four-dimensional
 harmonic oscillator, which gives here the squared-mass operator,
 the definite-metric-type subsidiary condition is adopted \cite{Takabayasi}.
 The space-time wave functions satisfying this condition
 is normalizable and leads to the desirable asymptotic behavior of
 electromagnetic form factors of hadrons. 
The eigenvalues of the squared-mass operator are given by
\begin{equation}
	M_N^2=M_0^2 + N\Omega,~~\Omega = 2(m_1+m_2) \sqrt{\frac{K}{\mu}},
\end{equation}
where $N = 2N_r+L$, $N_r~(L)$ being the radial  (orbital angular momentum) quantum number.
 This squared-mass formula gives linear Regge trajectories with the slope $\Omega^{-1}$,
 in accord with the well-known experimental fact.

\section{Electromagnetic currents for quark-antiquark meson systems}

The above Klein-Gordon-type equation is rewritten in terms of the quark and
 antiquark coordinates as
\begin{equation}
	2(m_1 + m_2)\left( \sum_{i=1}^2 \frac{-1}{2m_{i}}\frac{\partial^{2}}
	{\partial x_{i\mu}\partial x_{i}^{\mu}}-U(x_{1},x_{2}) \right)
	\Psi(x_{1},x_{2})_{\alpha}{}^{\beta} = 0.
\end{equation}
This equation is derived from either of  the actions
\begin{equation}
	S_\text{free}^{(\text{KG,S})} = \int d^4 x_1 \int d^4 x_2~~
	\mathcal{L}_\text{free}^{(\text{KG,S})}(\Psi, \partial_{1\mu}\Psi, \partial_{2\mu}\Psi)
\end{equation}
with the Klein-Gordon-like Lagrangian
\begin{equation}
	\mathcal{L}_{\rm{free}}^{\rm{(KG)}} = \overline{\Psi}(x_1,x_2)
	2(m_1 + m_2)\left(\sum_{i=1}^2 \frac{-1}{2m_{i}}
	\frac{\overleftarrow{\partial}}{\partial x_{i\mu}}
	\frac{\overrightarrow{\partial}}{\partial x_i^{\mu}}
	-U(x_{1},x_{2}) \right)\Psi(x_{1},x_{2})
\end{equation}
and the Schr\"{o}dinger-like Lagrangian
\begin{equation}
	\mathcal{L}_{\rm{free}}^{\rm{(S)}}=\overline{\Psi}(x_1,x_2)
	\left(\sum_{i=1}^2 \frac{-1}{2m_{i}}\frac{\overleftarrow{\partial}}
	{\partial x_{i\mu}}\frac{\overrightarrow{\partial}}{\partial x_i^{\mu}}
	-U(x_{1},x_{2}) \right)\Psi(x_{1},x_{2}),
\end{equation}
where the bilocal field $\Psi(x_{1},x_{2})$ has the dimension of bosons $[\mathrm{M}^1]$
 and fermions $[\mathrm{M}^{3/2}]$ for $\mathcal{L}_{\rm{free}}^{\rm{(KG)}}$ and
 $\mathcal{L}_{\rm{free}}^{\rm{(S)}}$, respectively, except for the dimension
 of internal wave functions $[\mathrm{M}^2]$.

The interaction of quark-antiquark meson systems with an electromagnetic field
 can be obtained \cite{Lipes} by the minimal substitutions
\begin{equation}
	\frac{\partial}{\partial x_{i}^{\mu}} \to 
	\frac{\partial}{\partial x_{i}^{\mu}} + ieQ_{i}A_{\mu}(x_{i})
\end{equation}
in the free Lagrangians $\mathcal{L}_{\rm{free}}^{\rm{(KG,S)}}$, in which
 the heuristic prescription by Feynman, Kislinger and Ravndal \cite{FKR}
 with some extension is adopted as the following replacements
\begin{equation}
	\frac{\overleftarrow{\partial}}{\partial x_{i\mu}}
	 \frac{\overrightarrow{\partial}}{\partial x_{i}^{\mu}}
	\to
	(1-g_M^{(i)})\frac{\overleftarrow{\partial}}{\partial x_{i\mu}}
	 \frac{\overrightarrow{\partial}}{\partial x_{i}^{\mu}}
	+ g_M^{(i)}\frac{\overleftarrow{\partial}}{\partial x_{i}^{\mu}}\gamma^{\mu} \gamma^{\nu}
	 \frac{\overrightarrow{\partial}}{\partial x_{i}^{\nu}},
\end{equation}
where $Q_{i}~(i=1,2)$ are the quark and antiquark charges in units of $e$ and
 $g_M^{(i)}$ are the parameters related to the anomalous magnetic moments
 of constituent quarks. Then the action for the electromagnetic interaction of
 meson systems is obtained, up to the first order of $e$, as
\begin{equation}
	S_\text{int}^{(\text{KG,S})} = \int d^4 x_1 \int d^4 x_2~
	\sum_{i=1}^2~j_{i}^{(\text{KG,S})\mu}(x_1,x_2) A_{\mu}(x_i)
	\equiv \int d^4 X J^{(\text{KG,S})\mu}(X) A_{\mu}(X)
\end{equation}
with the conserved currents
\begin{equation}
	j_i^{\rm{(KG)}\mu}(x_1, x_2) = 2(m_1+m_2)
	\left\langle \overline{\Psi}(x_1, x_2)\frac{-ieQ_i}{2m_i}
	\left(
	\frac{\overleftrightarrow{\partial}}{\partial x_{i\mu}}
	-ig_M^{(i)} \sigma^{\mu\nu} \left(
	\frac{\overrightarrow{\partial}}{\partial x_i^{\nu}}
	+ \frac{\overleftarrow{\partial}}{\partial x_i^{\nu}}
	\right)
	\right) \Psi(x_1, x_2)\right\rangle
\end{equation}
and
\begin{equation}
	j_i^{\rm{(S)}\mu}(x_1, x_2) = \frac{j_i^{\rm{(KG)}\mu}(x_1, x_2)}{2(m_1+m_2)}
\end{equation}
where $\langle \cdots \rangle$ means taking trace concerning the Dirac indices.
 The electric charges of meson systems are given by the diagonal elements
\begin{equation}
	Q_{\rm meson}^{(\text{KG,S})}
	 = \int d^3 X ~ \langle i|J^{(\text{KG,S})0}(X)|i \rangle.
\end{equation}
The respective features of the Klein-Gordon-like and Schr\"{o}dinger-like currents
 are as follows:
 
\begin{itemize}
 \item \textit{The Klein-Gordon-like current}.
The meson charge is given by
\begin{equation}
	Q_{\rm meson}^{(\text{KG})} = (Q_1+Q_2)e,
\end{equation}
which reproduces the physical one correctly.
 The currents $j_i^{\rm{(KG)}\mu}(x_1, x_2)$ have no absolute mass scales
 of constituent quarks. This would seem to be natural for light-quark systems
 from the viewpoint of QCD in the chiral limit.
 
 \item \textit{The Schr\"{o}dinger-like current}. 
In this case the meson charge becomes
\begin{equation}
	Q_{\rm meson}^{(\text{S})} = \frac{M}{m_1+m_2}(Q_1+Q_2)e,
\end{equation}
where $M$ is the meson mass. This expression does not generally coincide with
 the physical meson charge. However, it gives the correct charge in the heavy
 quark limit. This means that the Schr\"{o}dinger-like current is applicable
 to heavy-light and heavy-heavy systems.
If the meson masses for nonrelativistic quark systems are written as
\begin{equation}
	M_n = (m_1 + m_2) + \mathcal{E}_n
\end{equation}
with the $n$-th excitation energy $\mathcal{E}_n$, then the meson charge can be
 expressed as
\begin{equation}
	Q_{\rm meson}^{(\text{S})}
	= \left(1 + \frac{\mathcal{E}_n}{M_0} \right)(Q_1+Q_2)e,
\end{equation}
where $M_0$ is a mass of the ground-state meson. From this expression it is found
 that the applicability of the schr\"{o}dinger-like current is estimated by
 the ratio of the excitation energy to the ground-state mass.
The currents $j_i^{\rm{(S)}\mu}(x_1, x_2)$, unlike $j_i^{\rm{(KG)}\mu}(x_1, x_2)$,
 have the absolute mass scales of constituent quarks, which is a desirable feature
 in describing the nonrelativistic quark systems.
\end{itemize}

From the above considerations, it is concluded that the dimension of bilocal meson
 fields is bosonic for light-quark systems, while fermionic for heavy-light and
 heavy-heavy systems, except for the dimension of internal wave functions
 $[\mathrm{M}^2]$. Our recent studies of the pionic decays of excited heavy-light,
 charmed and charmed-strange, mesons support this conclusion for heavy-light
 systems \cite{Maeda}.

\section{Radiative decay widths of light-quark mesons}

The above effective electromagnetic interactions $S_\text{int}^{(\text{KG,S})}$
 describe systematically all the electromagnetic interactions of quark-antiquark
 meson systems. For the radiative decays of light-quark mesons the decay width
 is obtained, following the usual procedure with the Klein-Gordon-like interaction
 $S_\text{int}^{(\text{KG})}$, as
\begin{equation}
	\Gamma = \frac{1}{2J_i+1} \frac{|\bf{q}|}{8\pi M_i^2}
	\sum_{\rm{spin}}|\mathcal{M}_{fi}|^2,
\end{equation}
where $M_i~(J_i)$ are the mass (spin) of the initial-state meson and $|\bf q|$
 is the photon three-momentum. In the actual applications of the above formula
 to radiative decay widths of light-quark mesons the physical masses are used
 for initial- and final-state mesons, except for the pion.

Here we restrict ourselves to a discussion on the radiative decays of light-quark mesons
 only with nonstrange quarks. Then the present radiative decay model has two parameters,
 $\Omega$ and $g_M~(\equiv g_M^{(u)}=g_M^{(d)})$, the inverse of the Regge slope
 and the parameter related to the anomalous magnetic moment of $u$ and $d$ quarks.
 We take the value of $\Omega = 1.14~\rm{GeV}^2$ and determine a value of $g_M$
 so as to fit the experimental width for $\rho(770)^{\pm} \rightarrow \pi^{\pm}\gamma$.
 We also treat the pion mass as an additional parameter, which is determined
 in the following two ways:

\begin{itemize}
 \item \textit{Case A}.
Assuming that the pion lies on the spin-singlet, $b_1(1235)$--$\pi_2(1670)$,
 trajectory, the effective pion mass becomes 0.476 GeV. Calculating the numerical
 width for $\rho(770)^{\pm} \rightarrow \pi^{\pm}\gamma$ with this pion mass,
 $g_M = 1.51$ is obtained.
 
 \item \textit{Case B}.
In the case of the pion and the $\pi_2(1670)$ lying on the same trajectory with
 $\Omega = 1.14~\rm{GeV}^2$, the effective pion mass is 0.719 GeV. In this case
 the width for $\rho(770)^{\pm} \rightarrow \pi^{\pm}\gamma$ is calculated with
 this pion mass only for the exponential factor of the formula, while
 the physical pion mass for the other part. This treatment gives $g_M = 0.743$.
\end{itemize}

\begin{table}
\renewcommand{\arraystretch}{1.1}
\caption{Theoretical and experimental widths for radiative decays of light-quark mesons.
 The experimental values of the initial-state masses and radiative widths are taken
 from the Particle Data Group (PDG) \cite{PDG},
 unless otherwise noted. As for the $\rho^0 \gamma$ decay of the $f_1(1285)$,
 the VES-based value was obtained from the VES measurement of the branching fraction
 to $\rho^0 \gamma$ \cite{VES} and the PDG total width,
 while the CLAS-based one was computed using the CLAS measurement of the branching ratio
 $\Gamma (\rho^0 \gamma)/\Gamma (\eta \pi \pi)$ \cite{CLAS} and the PDG total width
 and branching fraction to $\eta \pi \pi$. The statistical and systematic errors
 of the respective measured values were added in quadrature.}
\label{tab1}
\begin{center}
\renewcommand{~}{\phantom{0}}
\begin{tabular}{lllll}
\hline \hline
	                                                &                      & \multicolumn{2}{c}{Theory (keV)} & \\
\cline{3-4}
	Decay process                                   & $M_i$ (MeV)          & Case A    & Case B    & Experiment (keV) \\
\hline
	$\rho(770)^{\pm} \rightarrow \pi^{\pm}\gamma$   & $~775.26 \pm 0.25$    & ~~68 (fit)& ~~68 (fit)& $~~~68 \pm 7$ \\
	$a_1(1260)^{\pm} \rightarrow \pi^{\pm}\gamma$   & $1230 \pm 40$        & ~464      & ~394      & $~~640 \pm 246$ \\
	$a_2(1320)^{\pm} \rightarrow \pi^{\pm}\gamma$   & $1318^{+0.5}_{-0.6}$ & ~340      & ~250      & $~~311 \pm 25$ \\
	$b_1(1235)^{\pm} \rightarrow \pi^{\pm}\gamma$   & $1229.5 \pm 3.2$     & ~~42      & ~~52      & $~~230 \pm 60$ \\
	$\pi_2(1670)^{\pm} \rightarrow \pi^{\pm}\gamma$ & $1672.2 \pm 3.0$     & ~~88      & ~150      & $~~181 \pm 29$ \\
	$f_1(1285) \rightarrow \rho(770)^0 \gamma$      & $1281.9 \pm 0.5$     & ~752      & ~508      & $~~636 \pm 240$ [VES] \\
	                                                &                      &           &           & $~~545 \pm 253$ [CLAS] \\
	                                                &                      &           &           & $~1203 \pm 331$ \\
\hline \hline
\end{tabular}
\end{center}
\end{table}

Numerical results of the radiative decay widths are shown in comparison with experiment
 in Table \ref{tab1}. The agreement with the measured widths for $\pi \gamma$ decays
 is satisfactory, though that for $\pi_2(1670)^{\pm} \rightarrow \pi^{\pm}\gamma$
 less so in Case A, except for $b_1(1235)^{\pm} \rightarrow \pi^{\pm}\gamma$.
 It should be noted that only the electric (convection) current, independent of
 $g_M$, contributes to the decay processes $b_1(1235)^{\pm} \rightarrow \pi^{\pm}\gamma$
 and $\pi_2(1670)^{\pm} \rightarrow \pi^{\pm}\gamma$.
 In addition, the width for the decay $f_1(1285) \rightarrow \rho(770)^0 \gamma$ can be
 calculated by just the same parameters, assuming that the $f_1(1285)$ is
 a pure $u\bar{u} + d\bar{d}$ state. The results are also shown in Table \ref{tab1},
 where the present model predictions are found to be in good agreement with
 the VES \cite{VES} and CLAS \cite{CLAS} measurements, not the PDG average value.

\section{Summary}

The dimension of bilocal meson fields should be bosonic for light-quark meson systems,
 while fermionic for heavy-light and heavy-heavy ones, except for the dimension
 of internal wave functions.
 The calculated results for the radiative decay widths of the $a_1(1260)$, $a_2(1320)$,
 $b_1(1235)$ and $\pi_2(1670)$ mesons to the pion are in fair agreement with experiment,
  aside from $b_1(1235)^{\pm} \rightarrow \pi^{\pm}\gamma$.
 For the decay $f_1(1285) \rightarrow \rho(770)^0 \gamma$ the present model strongly
 supports the experimental results of VES and CLAS Collaborations.

\end{document}